



\input lanlmac
\input amssym
\input epsf

\newcount\figno
\figno=0
\def\fig#1#2#3{
\par\begingroup\parindent=0pt\leftskip=1cm\rightskip=1cm\parindent=0pt
\baselineskip=13pt
\global\advance\figno by 1
\midinsert
\epsfxsize=#3
\centerline{\epsfbox{#2}}
\vskip 12pt
{\bf Fig. \the\figno:~~} #1 \par
\endinsert\endgroup\par
}
\def\figlabel#1{\xdef#1{\the\figno}}
\newdimen\tableauside\tableauside=1.0ex
\newdimen\tableaurule\tableaurule=0.4pt
\newdimen\tableaustep
\def\phantomhrule#1{\hbox{\vbox to0pt{\hrule height\tableaurule
width#1\vss}}}
\def\phantomvrule#1{\vbox{\hbox to0pt{\vrule width\tableaurule
height#1\hss}}}
\def\sqr{\vbox{%
  \phantomhrule\tableaustep

\hbox{\phantomvrule\tableaustep\kern\tableaustep\phantomvrule\tableaustep}%
  \hbox{\vbox{\phantomhrule\tableauside}\kern-\tableaurule}}}
\def\squares#1{\hbox{\count0=#1\noindent\loop\sqr
  \advance\count0 by-1 \ifnum\count0>0\repeat}}
\def\tableau#1{\vcenter{\offinterlineskip
  \tableaustep=\tableauside\advance\tableaustep by-\tableaurule
  \kern\normallineskip\hbox
    {\kern\normallineskip\vbox
      {\gettableau#1 0 }%
     \kern\normallineskip\kern\tableaurule}%
  \kern\normallineskip\kern\tableaurule}}
\def\gettableau#1 {\ifnum#1=0\let\next=\null\else
  \squares{#1}\let\next=\gettableau\fi\next}

\tableauside=1.0ex
\tableaurule=0.4pt


\def\Tr{{\rm Tr}}
\def\hf{{1\over 2}}

\def\o{\over}

\def\Si{\Sigma}
\def\b#1{\overline{#1}}

\def\bra{\langle}
\def\ket{\rangle}
\def\lf{\left}
\def\ri{\right}
\def\riya{\rightarrow}

\def\h#1{\widehat{#1}}

\def\Ga{\Gamma}
\def\al{\alpha}

\def\sitarel#1#2{\mathrel{\mathop{\kern0pt #1}\limits_{#2}}}

\def\cob{\delta}

\lref\OkuyamaZN{
  K.~Okuyama,
  ``N = 4 SYM on R x S(3) and pp-wave,''
  JHEP {\bf 0211}, 043 (2002)
  [arXiv:hep-th/0207067].
}
\lref\BergereAZ{
  M.~C.~Bergere,
  ``Correlation Functions of Complex Matrix Models,''
  arXiv:hep-th/0511019.
}
\lref\BeisertBB{
  N.~Beisert, C.~Kristjansen, J.~Plefka, G.~W.~Semenoff and M.~Staudacher,
  ``BMN correlators and operator mixing in N = 4 super Yang-Mills theory,''
  Nucl.\ Phys.\ B {\bf 650}, 125 (2003)
  [arXiv:hep-th/0208178].
}
\lref\BerensteinKK{
  D.~Berenstein,
  ``A toy model for the AdS/CFT correspondence,''
  JHEP {\bf 0407}, 018 (2004)
  [arXiv:hep-th/0403110].
}
\lref\CorleyZK{
  S.~Corley, A.~Jevicki and S.~Ramgoolam,
  ``Exact correlators of giant gravitons from dual N = 4 SYM theory,''
  Adv.\ Theor.\ Math.\ Phys.\  {\bf 5}, 809 (2002)
  [arXiv:hep-th/0111222].
}
\lref\PearsonZS{
  J.~Pearson, M.~Spradlin, D.~Vaman, H.~L.~Verlinde and A.~Volovich,
  ``Tracing the string: BMN correspondence at finite J**2/N,''
  JHEP {\bf 0305}, 022 (2003)
  [arXiv:hep-th/0210102].
}
\lref\Ginibre{
J. Ginibre,
``Statistical Emsembles of Complex, Quaternion and Real Matrices,''
J. Math. Phys. {\bf 6}, 440 (1965).
}
\lref\EynardDF{
  B.~Eynard and C.~Kristjansen,
  ``BMN correlators by loop equations,''
  JHEP {\bf 0210}, 027 (2002)
  [arXiv:hep-th/0209244].
}
\lref\ConstableHW{
  N.~R.~Constable, D.~Z.~Freedman, M.~Headrick, S.~Minwalla, L.~Motl, A.~Postnikov and W.~Skiba,
  ``PP-wave string interactions from perturbative Yang-Mills theory,''
  JHEP {\bf 0207}, 017 (2002)
  [arXiv:hep-th/0205089].
}
\lref\KristjansenBB{
  C.~Kristjansen, J.~Plefka, G.~W.~Semenoff and M.~Staudacher,
  ``A new double-scaling limit of N = 4 super Yang-Mills theory and PP-wave
  strings,''
  Nucl.\ Phys.\ B {\bf 643}, 3 (2002)
  [arXiv:hep-th/0205033].
}
\lref\CordesFC{
  S.~Cordes, G.~W.~Moore and S.~Ramgoolam,
  ``Lectures on 2-d Yang-Mills theory, equivariant cohomology and topological
  field theories,''
  Nucl.\ Phys.\ Proc.\ Suppl.\  {\bf 41}, 184 (1995)
  [arXiv:hep-th/9411210].
}
\lref\DouglasWY{
  M.~R.~Douglas,
  ``Conformal field theory techniques in large N Yang-Mills theory,''
  arXiv:hep-th/9311130.
}
\lref\BerensteinAA{
  D.~Berenstein,
  ``Large N BPS states and emergent quantum gravity,''
  arXiv:hep-th/0507203.
}
\lref\KinneyEJ{
  J.~Kinney, J.~Maldacena, S.~Minwalla and S.~Raju,
  ``An index for 4 dimensional super conformal theories,''
  arXiv:hep-th/0510251.
}
\lref\LinNB{
  H.~Lin, O.~Lunin and J.~Maldacena,
  ``Bubbling AdS space and 1/2 BPS geometries,''
  JHEP {\bf 0410}, 025 (2004)
  [arXiv:hep-th/0409174].
}
\lref\BerensteinJQ{
  D.~Berenstein, J.~M.~Maldacena and H.~Nastase,
  ``Strings in flat space and pp waves from N = 4 super Yang Mills,''
  JHEP {\bf 0204}, 013 (2002)
  [arXiv:hep-th/0202021].
}
\lref\CorleyMJ{
  S.~Corley and S.~Ramgoolam,
  ``Finite factorization equations and sum rules for BPS correlators in  N = 4
  SYM theory,''
  Nucl.\ Phys.\ B {\bf 641}, 131 (2002)
  [arXiv:hep-th/0205221].
}
\lref\DonosVM{
  A.~Donos, A.~Jevicki and J.~P.~Rodrigues,
  ``Matrix model maps in AdS/CFT,''
  arXiv:hep-th/0507124.
}
\lref\RodriguesEC{
  J.~P.~Rodrigues,
  ``Large N spectrum of two matrices in a harmonic potential and BMN
  energies,''
  arXiv:hep-th/0510244.
}
\lref\JevickiYI{
  A.~Jevicki,
  ``Nonperturbative collective field theory,''
  Nucl.\ Phys.\ B {\bf 376}, 75 (1992).
}
\lref\deMelloKochNQ{
  R.~de Mello Koch, A.~Jevicki and J.~P.~Rodrigues,
  ``Collective string field theory of matrix models in the BMN limit,''
  Int.\ J.\ Mod.\ Phys.\ A {\bf 19}, 1747 (2004)
  [arXiv:hep-th/0209155].
}
\lref\EynardDF{
  B.~Eynard and C.~Kristjansen,
  ``BMN correlators by loop equations,''
  JHEP {\bf 0210}, 027 (2002)
  [arXiv:hep-th/0209244].
}
\Title{             
                                             \vbox{
                                             \hbox{hep-th/0511064}}}
{\vbox{
\centerline{1/2 BPS Correlator and Free Fermion}
}}

\vskip .2in

\centerline{Kazumi Okuyama}

\vskip .2in

\centerline{Department of Physics and Astronomy, 
University of British Columbia} 
\centerline{Vancouver, BC, V6T 1Z1, Canada}
\centerline{\tt kazumi@phas.ubc.ca}
\vskip 3cm
\noindent

We propose that in the BMN limit
the effective interaction vertex in the 1/2 BPS sector
of ${\cal N}=4$ SYM is given by the Das-Jevicki-Sakita Hamiltonian.
We check for some examples that it reproduces the $1/N$ correction
to the correlation functions of $1/2$ BPS operators.

\Date{November 2005}

\vfill
\vfill

\newsec{Introduction}
In \refs{\CorleyZK,\BerensteinKK}, 
it was suggested that the 1/2 BPS sector of ${\cal N}=4$
SYM is described as a matrix quantum mechanics in a harmonic potential
(see also \refs{\DonosVM,\RodriguesEC}).
It is well-known that the Hilbert space 
of this system is represented by a free fermion system. 
In a recent paper \LinNB, the gravity dual of these 1/2 BPS 
states are constructed and they are indeed characterized by the
incompressible Fermi fluid in a two-dimensional plane.

The 1/2 BPS operators are given by the product of traces of
the scalar field $Z=X_1+iX_2$ in ${\cal N}=4$ SYM.
Their two-point functions 
do not receive quantum corrections, so they are given by
\eqn\twopt{
\lf\bra\prod_{i=1}^k\Tr Z^{J_i}(x)\prod_{j=1}^l\Tr \b{Z}^{K_j}(y)\ri
\ket_{{\cal N}=4~{\rm SYM}}={G_{\{J_i\}\{K_j\}}\o |x-y|^{2J}}
}
where $J$ is the total $U(1)$ charge
\eqn\Jsum{
J=\sum_{i=1}^kJ_i=\sum_{j=1}^lK_j,
}
and it is equal to the conformal dimension of the 1/2 BPS operator.
The non-trivial information is solely contained in the
numerical factor $G_{\{J_i\}\{K_j\}}$ in \twopt.
It was shown in \refs{\KristjansenBB,\ConstableHW}
that this factor is written as a complex Gaussian matrix integral
\eqn\gmat{
G_{\{J_i\}\{K_j\}}=\int[dZd\b{Z}]e^{-\Tr(Z\b{Z})}
\prod_{i=1}^k\Tr Z^{J_i}\prod_{j=1}^l\Tr \b{Z}^{K_j}.
}
It is further argued that in the BMN limit \BerensteinJQ
\eqn\BMN{
N,J\riya\infty~~{\rm with} ~~g_2={J^2\o N}~~{\rm fixed}
}
the non-planar diagrams survive in the computation
of $G_{\{J_i\}\{K_j\}}$ and it becomes
a non-trivial function of $g_2$.

In this short note, we try to connect the above two facts.
We argue that the BMN limit of $G_{\{J_i\}\{K_j\}}$
can be computed from the free fermion picture.
This paper is organized as follows. In section 2, we review the
Schur polynomial as the orthogonal basis of 1/2 BPS operators and
their relation to the free fermions.
In section 3, we propose that the BMN limit of two-point function can
be reproduced from the Das-Jevicki-Sakita Hamiltonian.
Section 4 is discussions. 

\newsec{Schur Polynomial and Free Fermion}
In this section, we review the relation between
 1/2 BPS operators and the Schur polynomials \refs{\CorleyZK,
\BerensteinKK,\CorleyMJ,\DonosVM}.
To write down the 1/2 BPS operators, it is useful to
introduce the free boson $\al_n$ obeying the standard
commutation relation
\eqn\boseCR{
[\al_n,\al_m]=n\cob_{n+m,0}.
}
Then we introduce the coherent state 
\eqn\cohstate{
|Z\ket=\exp\lf(\sum_{n=1}^\infty{1\o n}\Tr Z^n\al_{-n}\ri)|0\ket
}
which satisfies
\eqn\alnvstrZ{
\al_J|Z\ket=\Tr Z^J|Z\ket.
}
In other words, the oscillator $\al_J$ corresponds to a single trace
operator $\Tr Z^J$. In particular the mode number of oscillator
corresponds to the length of the trace, which in turn is identified as the
length of string via the spin chain picture.
More generally, the multi-trace operators correspond to the 
product of boson oscillators
\eqn\prodal{
\prod_{i=1}^k\al_{J_i}|Z\ket=\prod_{i=1}^k\Tr Z^{J_i}|Z\ket.
}

The basis of operators \prodal\ is not diagonal with respect to
the two-point function. The diagonal basis is obtained by fermionizing
the boson $\al_n$
\eqn\bosonize{
\al_n=\sum_{r\in {\Bbb Z}+\hf}c_{n-r}b_r
}
where $b_r$ and $c_r$ obey the anti-commutation relation
\eqn\fermiCR{
\{c_r,b_s\}=\cob_{r+s,0}.
}
For a given Young diagram $R$, we introduce the state $\bra R|$ as
\eqn\generalR{
\bra R|=\bra 0|\prod_{i=1}^sc_{h_i-i+1/2}\prod_{j=1}^sb_{v_j-j+1/2}
}
where $h_i$ are the row-lengths and $v_j$ are the column-lengths,
and $s$ is the number of boxes along the diagonal. 
For example, the diagram in the Fig. 1 corresponds to the state
\eqn\exstate{
\lf\bra\tableau{5 3 3 1}\ri|=\bra0|c_{9/2}c_{3/2}c_{1/2}b_{7/2}
b_{3/2}b_{1/2}.
}
\fig{Young Diagram and Fermions. The diagram is split into two parts by the
diagonal line (dashed line in the figure). The mode number $r$
of $c_r$ and $b_r$
corresponds to the (number of boxes)$+1/2$ in the direction 
shown by the arrows.
The number of boxes along the diagonal is $s$. ($s=3$ in this figure.)
}{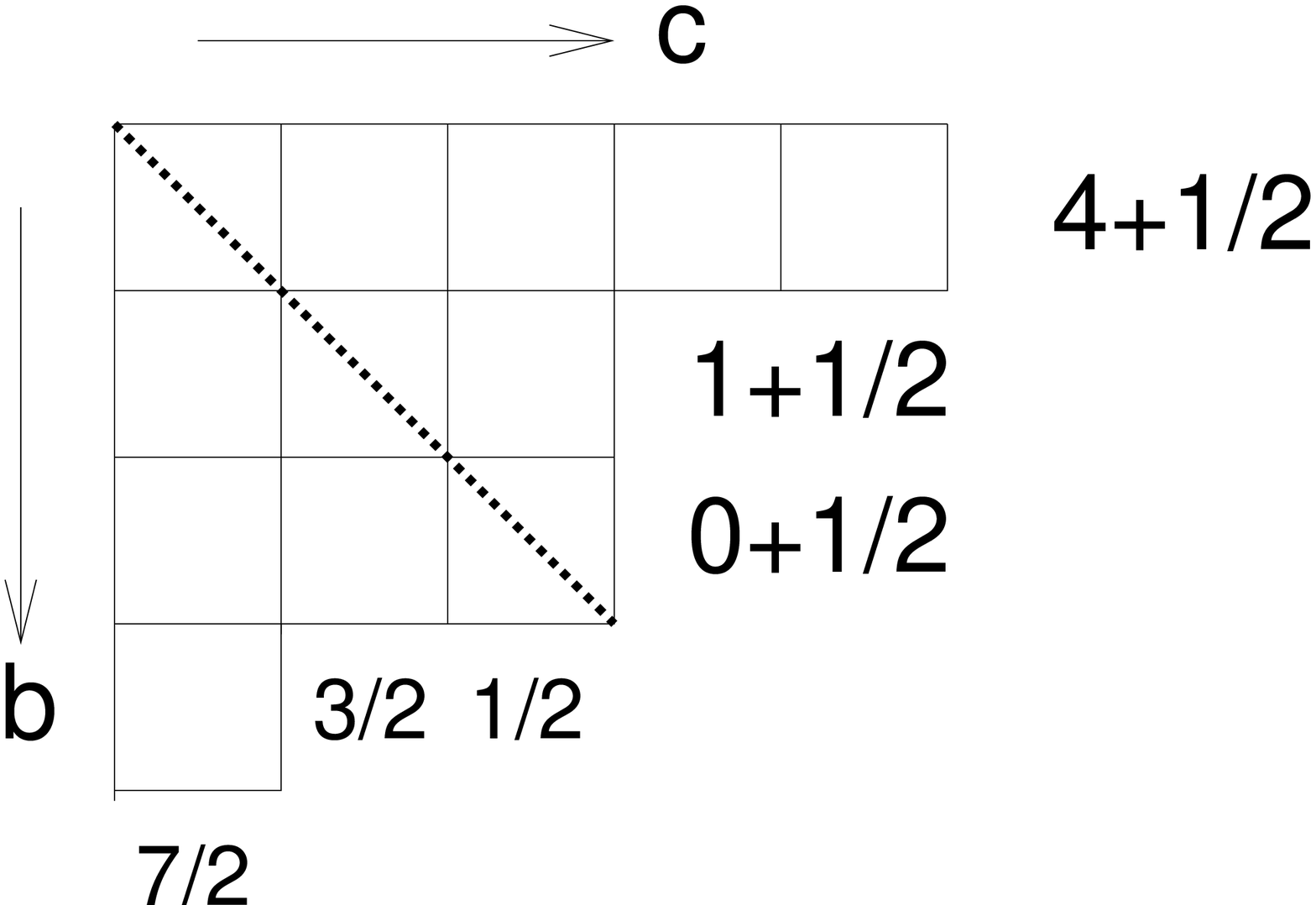}{5cm}
In terms of the state $\bra R|$ in \generalR, the Schur polynomial is defined by
\eqn\schurdef{
\bra R|Z\ket=S_R(Z).
}
This basis is diagonal with respect to the Gaussian measure
for the $N\times N$ complex matrix $Z$
\eqn\orthogo{
\int [dZd\b{Z}]e^{-\Tr(Z\b{Z})}S_R(Z)S_{R'}(\b{Z})={\cal N}_R\cob_{R,R'}.
}
For instance, at the level $L_0=2$ there are two states
\eqn\bose{
\bra\tableau{2}|=\bra0|c_{3/2}b_{1/2},\quad
\bra\tableau{1 1}|=\bra0|c_{1/2}b_{3/2}
}
and the corresponding Schur polynomial is given by
\eqn\schurex{
\bra\tableau{2}|Z\ket=\hf\Big[\Tr Z^2+(\Tr Z)^2\Big],\quad
\bra\tableau{1 1}|Z\ket=\hf\Big[\Tr Z^2-(\Tr Z)^2\Big].
}

In the computation of overlap integral \orthogo,
the following kernel naturally appears
\eqn\intVexact{
e^{{1\o 2N}{\cal V}_{\rm int}}\equiv
\int [dZd\b{Z}]e^{-\Tr(Z\b{Z})}|Z\ket\bra Z|.
}
Once we know the ``interaction vertex'' ${\cal V}_{\rm int}$ defined by
\intVexact, it is straightforward to compute the overlap integral
in the free boson picture
\eqn\alZint{
\bra0|\prod_{i=1}^k\al_{J_i}e^{{1\o2N}{\cal V}_{\rm int}}
\prod_{j=1}^{l}\al_{-K_j}|0\ket
=\int [dZd\b{Z}]e^{-\Tr(Z\b{Z})}
\prod_{i=1}^k\Tr Z^{J_i}\prod_{j=1}^{l}\Tr \b{Z}^{K_j}.
}

\newsec{${\cal V}_{\rm int}$ as Das-Jevicki-Sakita Hamiltonian}
Although it is known how to perform the Gaussian matrix integral
\Ginibre,
it is not so easy to evaluate the 
$Z$-integral \intVexact\ in a simple form 
and write down the interaction vertex ${\cal V}_{\rm int}$
in terms of oscillators $\al_n$.
However, we expect that ${\cal V}_{\rm int}$ simplifies in some particular
limit.
We propose that in the BMN limit ${\cal V}_{\rm int}$ can be replaced
by the Das-Jevicki-Sakita Hamiltonian ${\cal V}_3$ \JevickiYI\
\foot{A collective field theory in the BMN limit was considered in
\deMelloKochNQ.}
\eqn\Vthree{
{\cal V}_3=\sum_{n,m>0}\al_{-n}\al_{-m}\al_{n+m}
+\al_{-n-m}\al_n\al_m.
}
This is motivated by the intuition that string interaction
can be written as the splitting/joining and the length of strings is conserved
in the BMN limit (see Fig. 2).
This vertex can be thought of as the lightcone string field interaction
in the 1/2 BPS sector of ${\cal N}=4$ SYM.
\fig{This diagram represents the splitting and
joining of strings. In this process the $U(1)_J$ charge, or 
the total length of strings is conserved. This diagram comes with a factor
of $N^{\chi}=N^{-1}$.}{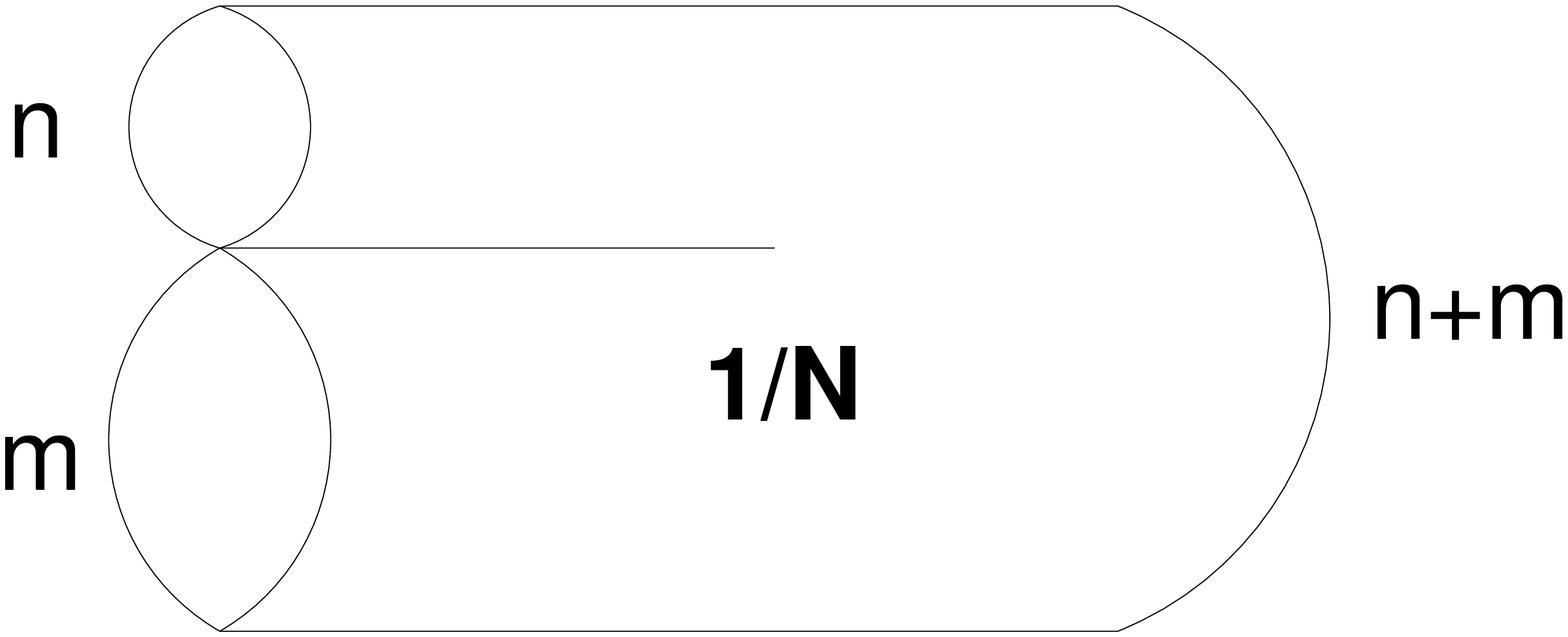}{5cm}
The important property of DJS Hamiltonian ${\cal V}_3$ is that
it is diagonal in the fermion basis ({\it i.e.} 
${\cal V}_3$ is a $W$-current) \DouglasWY 
\eqn\Vthbc{
{\cal V}_3=\sum_{r}r^2c_{-r}b_r.
}
We can easily see that ${\cal V}_3$ measures the square
of the mode number of fermions
\eqn\Vdiag{
[{\cal V}_3,c_r]=r^2c_r,\quad [{\cal V}_3,b_r]=-r^2b_r.
}
In particular, ${\cal V}_3$ is diagonal in the representation basis
and its eigenvalue is given by the second Casimir
\eqn\RRp{
\bra R|e^{{1\o2N}{\cal V}_3}|R'\ket
=e^{{1\o2N}C_2(R)}\cob_{R,R'}.
}
To compute the two-point function of states given by
the boson basis, we have to rewrite them in terms of the fermions:
\eqn\alnexp{\eqalign{
\prod_{i=1}^k\al_{-J_i}|0\ket
=&\sum_{S\subset\{1,\cdots,k\}}(-1)^{|S|+1}\sum_{0<r<J_S}
c_{-J+r}b_{-r}|0\ket\cr
+&\hf\sum_{A\cup B=\{1,\cdots,k\}}\sum_{S\subset A}\sum_{T\subset B}
(-1)^{|S|+|T|}\sum_{0<r<J_S}\sum_{0<s<J_T}
c_{-J_A+r}b_{-r}c_{-J_B+s}b_{-s}|0\ket
\cr
+&\cdots
}}
where $|S|$ denotes the number of elements in $S$, and
$J_S=\sum_{i\in S}J_i$. 
The first line in \alnexp\ is the state with one particle-hole pair,
and the second line is the state with two particle-hole pairs.
The dots denote the higher particle-hole pair states.

Below, we check our proposal \Vthree\ for two examples.
\subsec{Example 1: the $1\riya k$ Amplitude}
Let us consider the process that a single trace operator
splits into an operator with $k$ traces, {\it i.e.} the amplitude 
$G_{\{J\}\{J_1,\cdots,J_k\}}$.
We would like to show that this amplitude is written by
using the DJS Hamiltonian ${\cal V}_3$ in the free fermion picture
\eqn\onetokal{
\bra0|\al_Je^{{1\o2N}{\cal V}_3}\prod_{i=1}^k\al_{-J_i}|0\ket
=\sum_{S\subset\{1,\cdots,k\}}(-1)^{|S|+1}\sum_{0<r<J_S}
\exp\lf[{1\o2N}\Big((J-r)^2-r^2\Big)\ri].
}
Note that in this process only a one particle-hole pair
contributes. Therefore, we used the first line in \alnexp\
to rewrite the amplitude in terms of the free fermions.
On the right hand side of \onetokal, we used the relation \Vdiag.
In the BMN limit, the summation over $r$ is replaced by the integral over
$x=r/J$
\eqn\intonek{\eqalign{
\bra0|\al_Je^{{1\o2N}{\cal V}_3}\prod_{i=1}^k\al_{-J_i}|0\ket
&\simeq\sum_{S\subset\{1,\cdots,k\}}(-1)^{|S|+1}
J\int_0^{J_S/J}dx \exp\lf[{J^2\o2N}(1-2x)\ri] \cr
&={N\o J}\sum_{S\subset\{1,\cdots,k\}}(-1)^{|S|}
\exp\lf[{J\o2N}(J_{\b{S}}-J_S)\ri] \cr
&={N\o J}\prod_{i=1}^k\lf[\exp\lf({JJ_i\o 2N}\ri)-
\exp\lf(-{JJ_i\o 2N}\ri)\ri].
}}
Here $\b{S}=\{1,\cdots,k\}-S$ is the complement of $S$.
In the second equality, we used the relation $\sum_S(-1)^{|S|}=0$.
Finally, the BMN limit of this amplitude is given by
\eqn\summaryonek{
\lim_{N,J\riya\infty,{J^2\o N}=g_2}
\bra0|\al_Je^{{1\o2N}{\cal V}_3}\prod_{i=1}^k\al_{-J_i}|0\ket
={J\o g_2}\prod_{i=1}^k2\sinh\lf({g_2\o 2}\h{J}_i\ri)
}
where $\h{J}_i=J_i/J$.

On the other hand, the corresponding
matrix integral is known at finite $N$
\refs{\KristjansenBB,\ConstableHW,\CorleyMJ
\OkuyamaZN\BeisertBB\EynardDF{--}\BergereAZ}
\eqn\matonek{
\int[dZd\b{Z}]e^{-\Tr(Z\b{Z})}\Tr\b{Z}^{J}\prod_{i=1}^k\Tr Z^{J_i}
={1\o J+1}\sum_{S\subset\{1,\cdots,k\}}(-1)^{|S|}
{\Ga(N+J_{\b{S}}+1)\o
\Ga(N-J_S)}
}
In the BMN limit \matonek\ becomes 
\eqn\BMNlimonek{
\lim_{N,J\riya\infty,{J^2\o N}=g_2}\int[dZd\b{Z}]e^{-\Tr(Z\b{Z})}\Tr\b{Z}^{J}\prod_{i=1}^k\Tr Z^{J_i}
=
{JN^J\o g_2}\prod_{i=1}^k2\sinh\lf({g_2\o2}\h{J}_i\ri).
}
From \summaryonek\ and \BMNlimonek, 
one can see that
the DJS Hamiltonian correctly reproduces the matrix integral up to an overall
factor $N^J$.
The factor $N^J$ can be taken care of by modifying the identification as
\eqn\modid{
e^{{1\o2N}{\cal V}_{\rm int}}\simeq N^{L_0}e^{{1\o2N}{\cal V}_3}.
}

\subsec{Example 2: the $2\riya2$ Amplitude}

Next example is the $2\riya 2$ amplitude $G_{\{J_1,J_2\}\{K_1,K_2\}}$.
The free fermion computation is straightforward as in example 1.
In this case, both 
one particle-hole pair
and two particle-hole pairs contribute to the amplitude. So we need
the first and the second line \alnexp\ to rewrite the bosons into fermions.
Explicitly, the two-boson state is written in terms of fermions as
\eqn\alalbc{\eqalign{
\al_{-K_1}\al_{-K_2}|0\ket=&\lf(\sum_{0<r<K_1}+\sum_{0<r<K_2}-\sum_{0<r<J}\ri)
c_{-J+r}b_{-r}|0\ket \cr
&+\sum_{0<r<K_1}\sum_{0<s<K_2}c_{-K_1+r}b_{-r}c_{-K_2+s}b_{-s}|0\ket.
}}
After a similar calculation as in example 1, 
the BMN limit of the $2\riya2$ amplitude is 
found to be
\eqn\twoal{\eqalign{
&\lim_{N,J\riya\infty,{J^2\o N}=g_2}\bra0
|\al_{J_1}\al_{J_2}e^{{1\o 2N}{\cal V}_3}
\al_{-K_1}\al_{-K_2}|0\ket\cr
=&{J\o g_2}2^3\sinh\lf({g_2\o2}\h{J}_1\ri)
\sinh\lf({g_2\o2}\h{J}_2\h{K}_1\ri)\sinh\lf({g_2\o2}\h{J}_2\h{K}_2\ri).
}}
Here we assumed that $J_2={\rm max}\{J_i,K_j\},J_1={\rm min}\{J_i,K_j\}$.

On the other hand, the BMN limit of the matrix integral is \BeisertBB
\eqn\mattwo{\eqalign{
&\lim_{N,J\riya\infty,{J^2\o N}=g_2}\int[dZd\b{Z}]e^{-\Tr(Z\b{Z})}
\Tr\b{Z}^{J_1}\Tr \b{Z}^{J_2}\Tr Z^{K_1}\Tr Z^{K_2} \cr
&={JN^J\o g_2}2^3\sinh\lf({g_2\o2}\h{J}_1\ri)
\sinh\lf({g_2\o2}\h{J}_2\h{K}_1\ri)\sinh\lf({g_2\o2}\h{J}_2\h{K}_2\ri).
}}
Again, the two computations \twoal\ and \mattwo\ agree up to a factor $N^J$.

\newsec{Discussion}
We have checked for two examples that ${\cal V}_{\rm int}$ defined in 
\intVexact\ can be replaced by the DJS Hamiltonian ${\cal V}_3$ \Vthree\
in the BMN limit.
This agreement strongly suggests that the identification \modid\
holds for the general
correlator $G_{\{J_i\}\{K_j\}}$  \gmat.
It would be nice to find a
general proof.
In \PearsonZS, a similar interaction vertex $\Si$ was introduced in the 
string bit picture, and it was shown that it reproduces
the correct $g_2$ dependence. It would be interesting
to relate their vertex $\Si$ and ours ${\cal V}_3$.
It is well-known that the 
DJS Hamiltonian naturally appears in the two-dimensional Yang-Mills
theory \refs{\DouglasWY,\CordesFC}. It would be interesting to
find its relation to the BMN limit of 1/2 BPS sector (see \CorleyMJ\ for 
a discussion on the relation of 1/2 BPS correlators and 2d Yang-Mills).
Finally, it would be extremely interesting to find a useful description
of the 1/4 and 1/8 BPS states (see \refs{\BerensteinAA,\KinneyEJ} 
for some attempts).

\vskip 6mm
\noindent
{\bf Acknowledgment:} I would like to thank A. Jevicki and J. P. Rodrigues
for useful comments.
\listrefs
\bye